\documentclass[12pt]{iopart}

\usepackage{amssymb}
\usepackage{amsthm}
\usepackage{graphicx}
\usepackage{amsfonts}
\usepackage{epsfig}
\usepackage{hyperref}

\begin{document}

\title{Ultracold atoms in optical lattices induced by photonic crystals}
\author{Angela M. Barragan, Ian Mondragon-Shem and Herbert Vinck-Posada}
\address{Instituto de F\'{i}sica, Universidad de Antioquia, A.A. 1226, Medell\'{i}n, Colombia}
\ead{ambarra@gfif.udea.edu.co}

\begin{abstract}
We propose a way of generating optical lattices embedded in photonic crystals. By setting up extended modes in photonic crystals, ultracold atoms can be mounted in different types of field intensity distributions. This novel way of constructing optical lattices can be used to produce more elaborate periodic potentials by manufacturing appropriate geometries of photonic crystals. We exemplify this with a square lattice and comment on the possibility of using geometries with defects.
\end{abstract}

\maketitle

\section{Introduction}

\noindent Periodic arrangements are quite ubiquitous in various physical systems. From naturally occurring crystal structures in solid state systems, to artificially made periodic arrangements such as photonic crystals and optical lattices, both theoretical and experimental research in these systems has provided us with an understanding of fundamental physical phenomena, as well as with the possibility of achieving technological applications. There has been a growing interest in recent years in understanding the fundamental properties of photonic crystals \cite{Yamamoto,Abdel,Akahane,vignolini:045603,Badolato} and of optical lattices \cite{tuchman:130403, catani:140401, pepino:140405, mishmash:140403}, particularly concerning their possible applications in quantum optics and condensed matter systems.

Photonic crystals are periodic arrays of dielectric media in which light propagates linearly in various ways, depending on the geometry of the system. On the other hand, optical lattices are periodic arrays of electromagnetic field arising from the interference pattern of concurrent laser fields. In this case, ultracold atoms propagate nonlinearly in ways which depend also on the specific geometry of the system. 

Both systems share important properties, namely the band structure of their energy spectrum as well as the possibility of producing confined and extended modes of light in photonic crystals \cite{Joannopoulos} and of matter in optical lattices\cite{morsch:179}. However, there are important physical differences which render each system unique. Interaction between atoms in optical lattices has led to research into the nature of nonlinear matter waves in these systems, revealing a vast spectrum of nonlinear behaviour(see e.g. \cite{stanescu:053639,peterson:150406, wang:051604, wu:235107}). By contrast, light waves in photonic crystals obey Maxwell's equations and, hence, always satisfy the superposition principle.

It is of conceptual importance to understand how light and matter waves in these two systems are comparable, and to investigate the possibility of using the light fields in photonic crystals to control the matter waves in optical lattices.  Such a study can be carried out in two regimes: in the high-particle-density and coherent regime (in the case of photonic crystals, we mean photons, and in the case of optical lattices, we mean atoms), a description using a classical field in both systems is appropriate; on the other hand, in the low particle density regime, quantum phenomena arise, and the description of both systems will require a microscopic study (for photonic crystals see, \cite{Greentree}, and for optical lattices, see \cite{PhysRevLett.81.3108}). In the present paper, we address the first regime of high particle density. We focus on two interesting connected possibilities: first, one particular type of wave manipulation, namely the generation of confined modes produced by defects in both periodic structures; and second, the propagation of nonlinear matter waves in the presence extended light fields produced in photonic crystals. We will see that, although defects are just one example of wave manipulation that would be interesting to implement experimentally to create elaborate optical lattices, there are technical difficulties that have to be met before a successful realization.

The present paper is organized as follows: First, we review some important properies of defects in photonic crystals and in optical lattices. Second, we explain how an optical lattice can be induced by extended modes in photonic crystals. Finally, we exemplify this new way of obtaining optical lattices by using a photonic crystal with a two-dimensional square geometry. We conclude by suggesting the potential of using defect modes of photonic crystals to manipulate nonlinear matter waves embedded in them, and comment on the difficulties that must be resolved before realizing it experimentally.

\section{Periodic defect}

\subsection{Photonic crystals}

\noindent  Photonic crystals (PC) are periodic arrays of different dielectric materials whose dispersion relation presents bandgaps depending on specific structural parameters. This is their most important feature, because it allows us to control the flow of light through the material with low energy loss. 

The dielectric contrast of a photonic crystal is what produces a gap in the energy spectrum, for a given value of the propagation vector $\mathbf{k}$ of the light field that propagates in the structure. Periodicity allows us to use the Bloch theorem to obtain with ease the band diagram and electric and magnetic field modes. \cite{Joannopoulos}
 
By introducing defects, it is possible to confine a single mode of the electromagnetic field in a very narrow space of the lattice. Furthermore, we can introduce an active medium, such as a quantum dot or a two level atom, inside the defect, thus making it possible to analize radiation-matter interactions \cite{Yablonovitch}.

In order to obtain an analytical solution for the master equation, use is made of Maxwell's equations in free space in cartesian coordinates for a square array of dielectric rods embedded in air; only the linear term in the electric displacement is taken into account. The solution consists on diagonalizing the resulting matrix which is obtained when we expand the dielectric function and the electric field in a basis of plane waves and introducing them in the corresponding Maxwell equation. The dielectric function has the form

\begin{equation}
\epsilon(x,y)=\sum_{n,m}\beta_{n m}e^{-ib_{n}x}e^{-ib_{m}y},\label{die21}
\end{equation}

\noindent where the $\beta_{n m}$ are the expansion coefficients of the dielectric function. These contain the information about the geometrical distribution of the dielectric rods and have the form

\begin{figure}[h]
\centering
\includegraphics[scale=0.45]{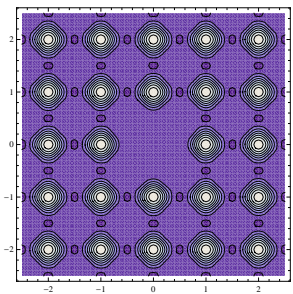}
\caption{Dielectric function of the PC supercell.}
\label{diel1}
\end{figure}

\begin{flushleft}
\begin{equation}
\hspace{-2cm}
\beta_{nm} = \sum_{x_{0},y_{0}=-N}^{N}{e^{i\frac{2\pi n}{L}x_{0}} e^{i\frac{2\pi m}{L}y_{0}}}\frac{(\varepsilon -1)}{A} \pi d^{2} \frac{J_{1}\left(\frac{1}{2}\sqrt{\left(\frac{2\pi n d}{L}\right)^{2}+\left(\frac{2\pi m d}{L}\right)^{2}}\right)}{\sqrt{\left(\frac{2\pi n d}{L}\right)^{2}+\left(\frac{2\pi m d}{L}\right)^{2}}}+ \delta_{n,0}\delta_{m,0}, \label{A1}
\end{equation}
\end{flushleft}

\noindent where $d$ is the diameter of the rods, $\varepsilon$ is the dielectric constant and $L=2N+1$ is the length of the supercell ($A=L^{2}$ is its area); $N$ is the number of layers around the defect. We have obtained  equation (\ref{A1}) analytically. The unitary cell for $N=2$ is shown in Fig.(\ref{diel1}). The electric field has the form

\begin{equation}
\textbf{E}(x,y)=\textbf{\^{z}}E_{z}(x,y)=\textbf{\^{z}}E_{p}(x,y)e^{-ik_{x}x}e^{-ik_{y}y},
\end{equation}

\noindent where $E_{p}(x,y)$ is the periodic function that takes into account the periodicity of the field in the $xy$ plane and the $b_{n}=b_{m}=\frac{2\pi n}{a}$ are the reciprocal vectors of the square array. The periodic function has the form

\begin{equation}
E_{p}(x,y)=\sum_{n}\sum_{m}\alpha_{nm}e^{-ib_{n}x} e^{-ib_{m}y}.\label{ezp}
\end{equation}

Introducing eq.(\ref{die21}) and eq.(\ref{ezp}) in Maxwell's equations we obtain the resulting eigenvalue equation

\begin{equation}
\alpha_{nm}\left[\left(\frac{2\pi n}{a}+k_{x}\right)^{2}+\left(\frac{2\pi m}{a}+k_{y}\right)^{2}\right]=k^{2}\sum_{p,q}{\alpha_{nm}\beta_{n-p,m-q}},\label{auto2}
\end{equation}

\noindent where the $\alpha_{nm}$ are the expansion coefficients of the electric field. We can note that this is a generalized eigenvalue problem whose eigenfunctions correspond to the expansion coefficients of the electric field with eigenvalues that correspond to the allowed energies for a fixed pair $k_{x}$ and $k_{y}$. With this solution, we construct the band diagram of the PC with a defect and we analyze the effect of introducing it in the crystal.

\subsection{Optical lattices}

\noindent Ultracold atoms in optical lattices are described in three different regimes depending on the particle density in the system \cite{morsch:179}. We focus on the limit of high-particle-density, in which the system remains in a macroscopic superfluid state. In this limit, we can make use of a classical field to describe the macroscopic coherent state of the atoms in the optical lattice, which is reminiscent of the classical field that describes light in photonic crystals. 

However, as has been noted, there is a marked difference between both systems: in the case of ultracold atoms, the waves behave nonlinearly, due to the interparticle interaction, whereas light always propagates linearly. This nonlinear description of the atomic case can be modeled using the well-known Gross-Pitaevskii (GP) equation:
\begin{equation}
\left[-\frac{\hbar^2\nabla^2}{2m}+V(\mathbf{r})+g \vert \psi(\mathbf{r}) \vert^2\right]\psi(\mathbf{r})=\mu \psi(\mathbf{r}),
\end{equation}
\noindent where $g$ denotes the nonlinear coupling. This equation can be derived using a Hartree-type ansatz to minimize the energy of the system, asuming $s-$wave scattering between the atoms. The wavefunction $\psi$ describes a macroscopic coherent state. In the case of optical lattices, the periodic potential is given by
\begin{equation}
V(\mathbf{r})=V_0 (\sin^2 kx+\sin^2 ky),
\end{equation}
\noindent where $V_0$ is proportional to the intensity of the laser light. Such a periodic potential can be generated by setting up standing waves with two lasers at perpendicular angles. The two-dimensional nature of this optical lattice can be achieved by setting up a tightly confining field in the third perpendicular direction, thus eliminating to a large extent the corresponding degree of freedom. 

The GP equation was numerically integrated with a procedure consisting of two stages. First, we perform a pseudospectral interpolation of the GP equation without the nonlinearity, i.e. of the Schr$\ddot{o}$dinger single-particle equation. Since the usual Bloch argument concerning periodic lattices in solid state physics applies also in optical lattices, we make use of the Bloch form of the wave function. Once we obtain the solution in $k$-space we reconstruct the spatial wave function as a superposition of plane waves, which amounts to making the plane-wave expansion that what was done in the photonic crystal case.

\begin{figure}[h]
\centering
\includegraphics[scale=0.42]{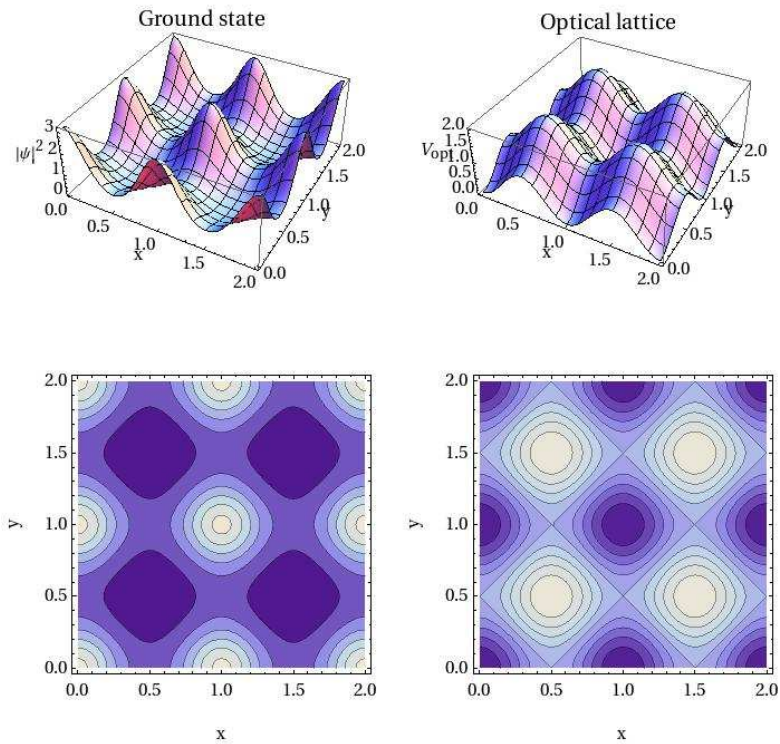}
\caption{\textbf{Left plots:} GP ground state of an optical lattice. \textbf{Right plots:} Optical lattice potential. }
\label{density}
\end{figure}

Afterwards, we propagate this single-particle wavefunction in imaginary time, using the GP equation. The evolution in imaginary time generates a superposition of eigenstates of the GP equation with exponentially decaying coefficients. By using this procedure, the term in the superposition with the lowest energy, i.e. the ground-state, will be the least suppressed in the evolution and, thus, will be predominant after some time has passed. In this way we will have obtained the GP ground state. It is important to notice that, because the GP equation is determined by the spatial density of the system, we need to normalize the wave function continuously in the time evolution. Otherwise, the system will not evolve with the correct effective GP Hamiltonian.

Using this same procedure, we can introduce a Gaussian defect in the optical lattice. Such defects have been studied in one-dimensional optical lattices, and have been shown to exhibit transport properties with dynanic solitonic behaviour due mainly to the nonlinearity inherent in the system \cite{brazhnyi}. However, here we focus mainly on the ground state of the system.

It should be noted that there is an important difference between defects in optical lattices and in photonic crystals concerning confined states. In experimental setups in optical lattices, it is the ground state which can be achieved by lowering the temperature of the system. As a result, in the band diagram of the optical lattice, we always choose the point of lowest energy. This contrasts to what is done in photonic crystals, where the interest focuses on confined modes that have an energy in the bandgap of the spectrum. In this case, it is not the lowest energy state which is chosen in order to confine light waves, but rather a higher energy state.

\section{Optical lattices induced by photonic crystals}

\noindent As we have discussed previously, optical lattices are stationary periodic light fields. Such lattices are usually generated through the use of interfering laser fields. However, it is not necessary to use laser fields. What matters is that we have a light field that couples to the atoms we are using through an effective external dipole potential. Such a dipole potential is proportional to the intensity as
\begin{equation}
U_{dip}(\mathbf{r})\sim I(\mathbf{r})
\end{equation}
where $I(\mathbf{r})$ denotes the field intensity. In particular, photonic crystals are suitable structures in which various types of distributions can be generated. As we reviewed in the previous section, by a suitable combination of geometry, it is possible to mold the field distribution by making use of, for example, defects. If we could take advantage of the vast set of intensity distributions in photonic crystals, we could achieve much more exotic behaviour than what has been achieved up until now in conventional optical lattices. 

\section{Results and discussion}

\noindent 
Before discussing an example of an optical lattice induced by a photonic crystal, let us first show some results of both systems independently, in which we can witness the use of defects as important means of generating more elaborate light fields in photonic crystals, and of studying nonlinear behaviour in optical lattices.

\begin{figure}[h]
\centering
\includegraphics[scale=0.45]{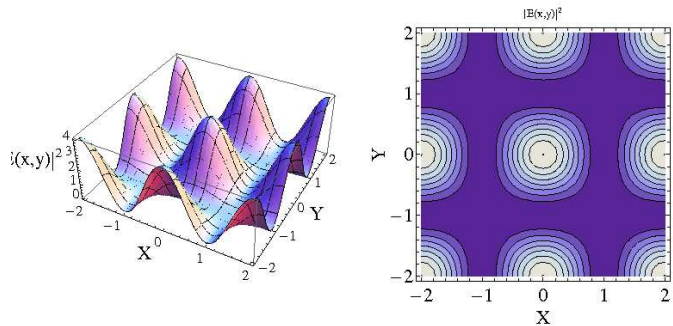}
\caption{Electric field of an extended mode in a PC. }
\label{elecfield}
\end{figure}

For the PC case, we have taken a finite basis of plane waves to expand eq.(\ref{auto2}) and to obtain a set of linear equations. After diagonalizing the resulting coefficient matrix, we have obtained the eigenfunctions corresponding to the electric field modes whose spatial distribution is shown in Fig. (\ref{elecfield}).

Point defects in PC confine light modes with a definite energy in a very narrow space. This fact can be used to construct cavities with a very high quality factor and obtain a strong coupling regime wich is important, for example, in the study of polaritonic systems. Furthermore, in these systems it is important to have a single mode of the field for simulating certain theoretical models in quantum optics such as the Jaynes-Cummings model.

\begin{figure}[h]
\centering
\includegraphics[scale=0.43]{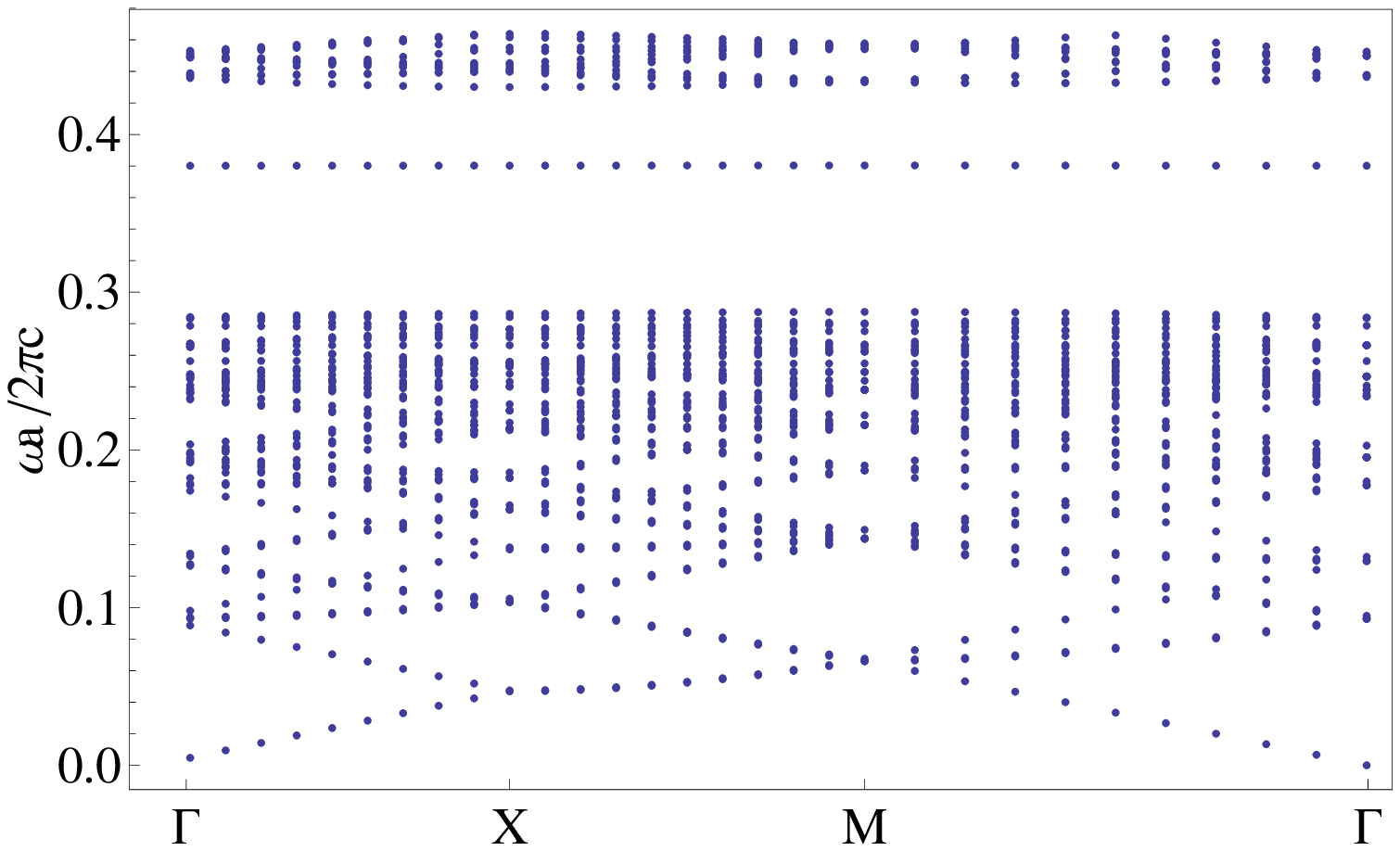}
\includegraphics[scale=0.5]{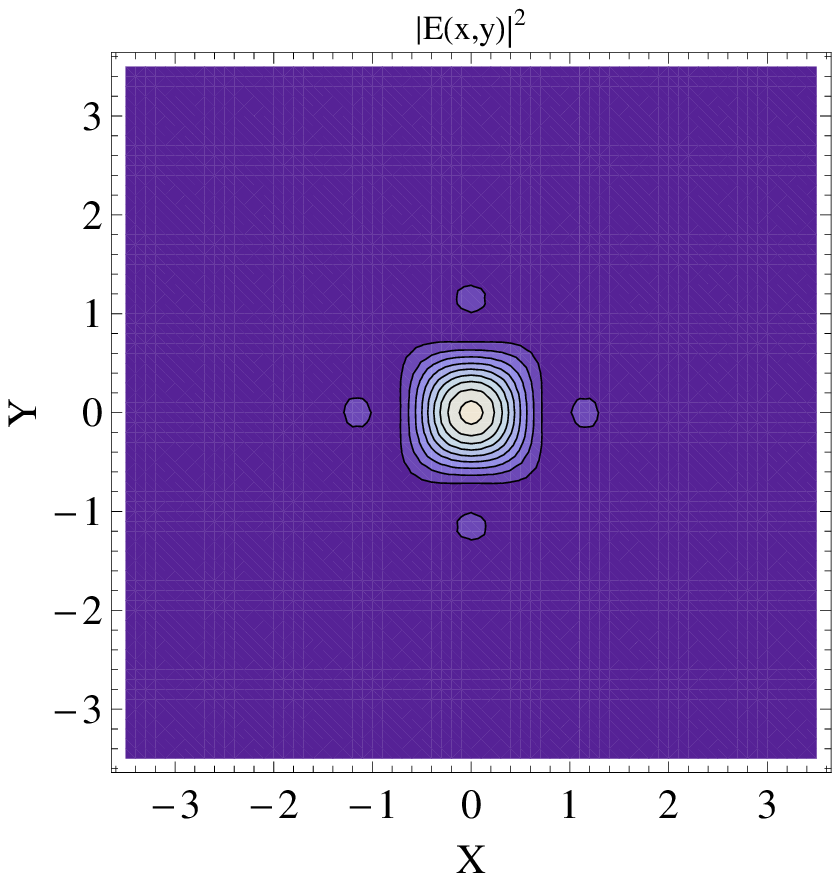}
\caption{\textbf{Left:} Band diagram of the photonic crystal with one defect per supercell. \textbf{Right:} Electric field pattern of the defective mode.}
\label{band1}
\end{figure}

We have obtained the TM modes of the electric field because the square geometry only presents bandgap for such modes. These are precisely the modes which can be confined when we introduce the defect in the PC, modifying their band diagram and allowing modes into the gap. The band diagram is shown in Fig.(\ref{band1}).

\begin{figure}[h]
\centering
\includegraphics[scale=0.33]{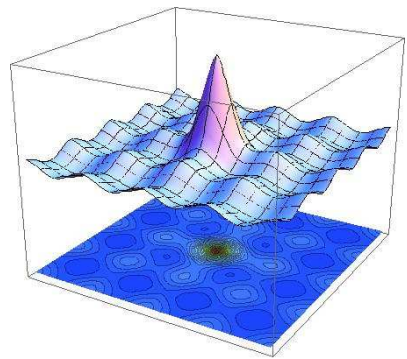}
\caption{GP ground state with a gaussian defect.}
\label{ground}
\end{figure}

It can be noticed that the field is highly, although not totally, concentrated in the defect region. The reason the field is largely concentrated inside the defect is due to the fact that point defects introduce defective modes in the bandgap. However, there is a region outside the defect where the intensity is not zero. This is due to the evanescent nature of the light in point defects that do not have quality factor high enough as to confine the field completely.

On the other hand, for the optical lattice case, we can see the resulting ground state in Fig.(\ref{ground}). This can be compared to the ground state without defect, shown in Fig. (\ref{density}). The localization effect of the wave function comes out of the minimization of the average energy of the system. Since the gaussian defect serves as a deeper quantum well than other parts of the optical lattice, it becomes more energetically favourable to occupy it, even though this effect competes with the increased positive interaction between the atoms when they are brought close together. 

The manipulations we show here of both light and matter waves are interesting on their own right. The molding of both types of waves through the use of appropriate geometries seems to offer a large number of possibilities. In particular, it would be interesting if we could use the molding of light in photonic crystals in order to further accomplish molding of matter waves. We now discuss a simplified example of this possibility using a system with perfect square periodicity for proof of principle. However, we do not introduce defects, or any special geometry, in this example because there is a technical difficulty in photonic crystals that must first be resolved before taking advantage of the actual molding of light to manipulate atoms. We will discuss this difficulty shortly.

\begin{figure}[h]
\centering
\includegraphics[scale=0.5]{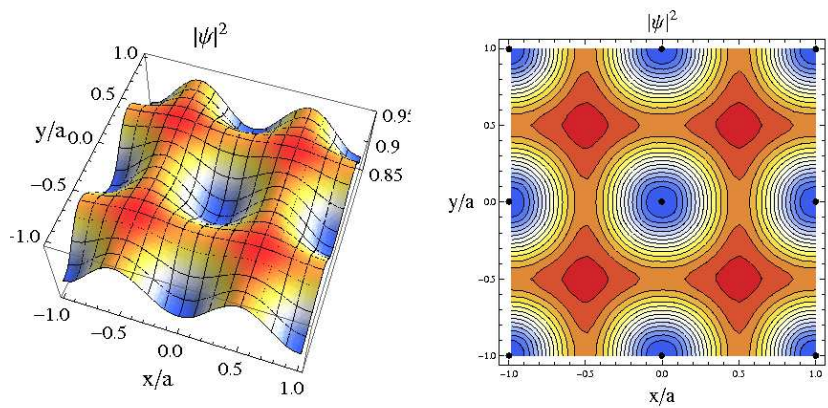}
\caption{Ground state for atoms in a OL created by one allowed mode in a PC. The black dots in the contour plot represent the dielectric posts of the photonic crystal.}
\label{ground2}
\end{figure}

In the following example, we have considered a square photonic crystal which has very thin dielectric posts (of the order of $0.01a$, with $a$ being the lattice constant), so that the cold atoms we use can be affected by the the periodic optical potential without being significantly affected by the material obstacles. We chose an extended mode to illustrate a simple example in which the potential is perfectly periodic. We show this in Fig.(\ref{elecfield}): the electric field of the extended mode is mainly concentrated at the dielectric posts, further inhibiting the atoms to come into close proximity of the dielectric posts. The Gross-Pitaevskii ground state was computed using this potential, and the resulting distribution is shown in Fig. (\ref{ground2}).

This distribution is markedly different from that of the usual optical lattice. The new effective potential is such that each potential minimum is not completely separated by a tunneling barrier from the neighboring minima but, instead, the potential tends to be flat. This facilitates the motion of the atoms between minima of the optical potential. Because of this, we see that the height of the wave function does not change significantly in the region between the posts, which contrasts with the ground state found in Fig.(\ref{density}). The solution shown in Fig.(\ref{ground2}) exhibits the basic effect of having an extended mode of the PC as an optical lattice. In an actual experimental setup it would be necessary to have an external parabolic potential to provide a confinement for the atoms; otherwise they would just escape from the photonic crystal.

We must address the deterrents which might make it difficult to use these types of optical lattices. The most important one is that if we use, for example, periodic defects in a photonic crystal to induce an optical lattice, we must take into account that the kinds of modes we can produce have some field minima inside the dielectric regions, which would promote collisions of atoms with the posts. Although this is not an actual disadvantage, it does require a careful theoretical study in order to quantify such effects. Furthermore, even if we can manage to have the light intensity concentrate at the posts, there will inevitably be some residual scattering of the atoms from the posts. This would make it difficult to model the system theoretically, as opposed to relatively simple hamiltonians that arise in the standard optical lattices. Finally, there is the experimental difficulty of mounting atoms to the photonic crystal and of performing, for example, time-of-flight measurements to probe the state of the system. However, we feel these experimental difficulties, although important, could be eventually resolved through suitable modifications of the usual experimental setups for optical lattices.

\section{Conclusions}

\noindent 

In the present work we have explored the possibility of using photonic crystals to induce an optical lattice. By using geometrical constructs such as defects, it is possible to manipulate and mold the intensity distribution of light waves in photonic crystals and the nonlinear dynamics of matter waves in optical lattices. We have argued that a suitable combination of both will provide a new source of many-body quantum behaviour.
Furthermore, using an extended mode in photonic crystals in order to generate the periodic potential for an optical lattice could be more favourable in reducing costs and in exploring new geometries which cannot be obtained easily with the conventional lasers used today. Thus, there is a practical, besides fundamental, justification for considering such new types of optical lattices.

\section*{Acknowledgments}
The authors acknowledge financial support from  the CODI-Universidad de Antioquia. We aknowledge J. P. Vasco, B. Rodriguez and P. Soares Guimaraes for enlightening discussions. 


\end{document}